\documentclass[fleqn,usenatbib]{mnras}

\usepackage{newtxtext,newtxmath}

\usepackage[T1]{fontenc}

\DeclareRobustCommand{\VAN}[3]{#2}
\let\VANthebibliography\thebibliography
\def\thebibliography{\DeclareRobustCommand{\VAN}[3]{##3}\VANthebibliography}

\usepackage{graphicx}	
\usepackage{amsmath}	
\usepackage{wasysym}
\usepackage{threeparttablex}
\usepackage{lscape} 
\usepackage{geometry}
\usepackage{longtable}
\usepackage[dvipsnames]{xcolor}
\usepackage{float}

\title[A rocky exoplanet classification method]{A rocky exoplanet classification method and its application to calculating surface pressure and surface temperature}

\author[S.R.N McIntyre et al.]{
Sarah R.N. McIntyre,$^{1,2}$\thanks{E-mail: sarah.mcintyre@anu.edu.au}
Penelope L. King,$^{2}$
and Franklin P. Mills$^{3,4}$
\\
$^{1}$Research School of Astronomy and Astrophysics, Australian National University, Canberra, ACT 2611, Australia\\
$^{2}$Research School of Earth Sciences, Australian National University, Canberra, ACT 2601, Australia\\
$^{3}$Fenner School of Environment and Society, Australian National University, Canberra, ACT 2601, Australia \\
$^{4}$Space Science Institute, Boulder, CO 80301, USA
}

\date{Accepted XXX. Received YYY; in original form ZZZ}

\pubyear{2023}

\begin{document}
\label{firstpage}
\pagerange{\pageref{firstpage}--\pageref{lastpage}}
\maketitle

\begin{abstract}
With over 5,000 exoplanets currently detected, there is a need for a primary classification method to prioritise candidates for biosignature observations. Here, we develop a classification method to categorise rocky exoplanets based on their closest solar system analogue using available data of observed stellar and planetary features, masses, and radii, to model non-thermal atmospheric escape, thermal atmospheric escape, and stellar irradiation boundaries. Applying this classification method to the 720 rocky exoplanets in our sample with uncertainties in planetary masses, radii, stellar temperatures, and fluxes propagated via a Monte Carlo model indicates that 22\% $\pm$ 8\% are Mercury analogues, 39\% $\pm$ 4\% are Mars analogues, 11\% $\pm$ 1\% are Venus analogues, 2\% $\pm$ 1\% are Earth analogues, and 26\% $\pm$ 12\% are without a known planetary counterpart in our solar system. Extrapolating to conditions on LHS 3844b and GJ 1252b, our classification method gives results reasonably consistent with current observations. Subsequently, to demonstrate the functionality of this classification method, we plot our catalogued sample of exoplanets on an adjusted surface pressure versus temperature phase diagram, presenting more realistic estimates of the potential surface phases (gas, liquid or ice). Our new classification method could help target selection for future exoplanet characterisation missions.
\end{abstract}




\section{Introduction}
Over the past 28 years, astronomers have observed over 5,000\footnote{\url{https://exoplanetarchive.ipac.caltech.edu/} (Accessed 18 December 2022) \label{nasa0}} extrasolar planets, providing us with basic information regarding their orbits, masses, and radii. Research on exoplanets is currently focused on determining which of these worlds may be habitable; for example, rocky bodies (with sufficient gravity to support an atmosphere) orbiting their host star at a distance where stellar insolation flux is suitable for the existence of liquid water on their surface \citep{kaltenegger2017characterize}. Due to the inability to conduct in-situ exploration, near-term studies to further characterise exoplanets will focus on remote detection of their atmospheres and spectral observations of possible biosignatures \citep{schwieterman2018exoplanet}. However, the significant observing time required to characterise rocky exoplanets limits the number of targets where we can conduct such extensive observations. 

Defining how potential atmospheric biosignatures vary under different conditions is important when characterising exoplanets \citep{schwieterman2018exoplanet}. Two significant parameters when considering the climatic conditions of an exoplanet are surface pressure and surface temperature \citep{keles2018effect}. Water’s stability on a planetary surface as a liquid depends on both the surface temperature and pressure \citep{seager2013exoplanet}. While the freezing point of liquid water is not strongly dependent on surface pressure, the boiling point is significantly affected by it \citep{vladilo2013habitable}. Furthermore, at surface pressures below the triple point, liquid water is not stable at any temperature. Thus, reliable estimates of the surface pressure and temperature are essential for characterising the environment and habitability of an exoplanet. Research suggests that a rise in surface pressure could lead to a rise in temperature due to the greenhouse effect \citep{kopparapu2014habitable}. However, a high surface pressure can enhance cooling through increased Rayleigh scattering \citep{kasting1993habitable,keles2018effect}, Mie scattering \citep{kitzmann2010clouds}, or reflection due to clouds \citep{marley2013clouds}. Additionally, exoplanet general circulation model (GDM) simulations and solar system observations suggest that high surface pressures could increase the latitudinal heat transport, cancelling seasonal variations in the planet’s surface temperature and resulting in smaller global temperature variations \citep{bullock1996stability,hansen2010global,leovy2001weather,trenberth2001estimates,vladilo2013habitable}. 

Despite the importance of surface pressure, current proposed methods for its measurement, using remote-sensing techniques, are challenging and may not fall within the wavelength cut-off for the James Webb Space Telescope \citep{chamberlain2013ground,crow2011views,gardner2006james,kasting2009exoplanet,misra2014using}. Three-dimensional general circulation models (3D GCMs) are beginning to provide insights into exoplanet atmospheres \citep{boutle2017exploring,del2019habitable,galuzzo2021three,lewis2022temperature,turbet2016habitability,turbet2018modeling,way2018climates,wolf2017assessing}. While there are a significant number of 3D GCM exoplanet simulations published given the lack of observational data so far, such models are computationally intensive and not generally accessible by the broader scientific community, limiting the number of simulations conducted to date \citep{del2019albedos}. Furthermore, many 3D GCM simulations of rocky exoplanets model ``Earth-like'' atmospheres, assuming $\sim$1 bar of N$_{2}$ as the predominant component of the atmosphere. 

An initial estimate of an exoplanet’s surface pressure (${P}_{surf}$) can be obtained from a simple model based on hydrostatic equilibrium \citep[e.g.][]{hall2020ensemble,kippenhahn1990stellar,kopparapu2014habitable,mordasini2012characterization,silva2017quantitative}, using available observational data on an exoplanet's mass (${M}_{p}$) and radius (${R}_{p}$):
\begin{equation} 
\label{eq:1} 
\frac{P_{surf}}{P_{\oplus}} = {\left(\frac{M_p}{M_\oplus}\right)}^{2} {\left(\frac{R_\oplus}{R_p}\right)}^{4}
\end{equation}

\noindent where $P_{\oplus}$, $M_\oplus$ and $R_\oplus$ are the surface pressure, mass, and radius of Earth, respectively. For many exoplanets, the radius or the mass is unknown, resulting in the publication of several mass-radius relations dependent on the planet’s type, allowing us to calculate the missing measurement \citep[e.g.][]{chen2016probabilistic,nikouravan2021estimating,otegi2020revisited,seager2007mass,swift2011mass,turbet2020revised,weiss2014mass,zeng2016mass}. Here, we follow the NASA exoplanet database\footnote{\url{https://exoplanetarchive.ipac.caltech.edu/} (Accessed 18 December 2022) \label{nasa2}} and use the \citet{chen2016probabilistic} M-R relationship to fill in the missing parameter:
\begin{equation} 
\label{eq:2} 
{R_p} \sim {{M_p}^\mathrm{0.279 \pm 0.009}} 
\end{equation}

Using the relation from Equation \ref{eq:2}, the surface pressure in Equation \ref{eq:1} can be written as:
\begin{equation} 
\label{eq:4} 
\frac{P_{surf}}{P_{\oplus}} = {\left(\frac{R_p}{R_\oplus}\right)}^{3.168 \pm 0.232}
\end{equation}

This Earth normalisation significantly limits the range of possible surface pressure values (Equations \ref{eq:5} and \ref{eq:6}), as evident when calculating the maximum and minimum ${P}_{surf}$ using Equation \ref{eq:4} and taking the upper radius limit for rocky exoplanets as $1.23 R_\oplus$, from the \cite{chen2016probabilistic} definition of the boundary between terrestrial and Jovian planets, and a lower radius limit of $0.3 R_\oplus$, which corresponds to the size of the smallest exoplanet discovered around a main-sequence star – Kepler 37b \citep{haghighipour2015kepler}.

\begin{equation} 
\label{eq:5} 
P_{surf max} = 1.014{\left(1.23\right)}^{3.168} = 1.95 bar
\end{equation}
\begin{equation} 
\label{eq:6} 
P_{surf min} = 1.014{\left(0.3\right)}^{3.168} = 0.02 bar
\end{equation}

The benefit of Equation \ref{eq:1} is that it only requires an exoplanet's radius or mass value. However, we obtain an Earth-centric estimate within the $0.02-1.95$ bar range because we use Earth's surface pressure and radius in our calculations. Our solar system contains four rocky planets, each with a unique surface, atmosphere, structure, and evolution. Extrasolar rocky planets will likely display a similarly wide variety of surface characteristics and interior compositions. While the radius range of $0.3-1.23 R_\oplus$ covers all four rocky planets in the solar system \citep{chen2016probabilistic,haghighipour2015kepler}, the resulting surface pressure range only encompasses Earth's. This simple model excludes the 5$\times10^{-15}-92$ bar range of surface pressure measurements observed on Mercury, Mars, and Venus \citep{rasool1966atmosphere,seiff1985models}. 

Alternative methods, such as comparing to a more appropriate solar system analogue, should be considered rather than continuing with the current approach of assigning Earth-like characteristics to all rocky exoplanets. A similar approach has been used for modelling the atmospheric chemistry and climate of Venus-like exoplanets \citep{kane2019venus,schaefer2011atmospheric,way2020venusian}. Furthermore, the scientific community has been studying, observing, and probing the planets of the solar system with multiple satellites, in situ missions, and remote sensing observations using ground- and space-based telescopes \citep{groller2018maven,jakosky2015mars,marcq2018composition,mcclintock2007mercury,mcnutt2010messenger,mills2006oxygen,von1979venus,withers2015comparison}. Thus, we have significant knowledge of the solar system planets' atmospheric profiles and compositions. We have values for the surface pressures of Mercury, Mars, and Venus and could use these planets' respective pressure and radii values to normalise Equation \ref{eq:1} more appropriately, provided we can classify which exoplanets are likely to be Mercury, Mars, or Venus analogues. 

Previous classification schemes from \citet{forget2014possible} and \citet{wordsworth2022atmospheres} suggest that climates on terrestrial exoplanets should depend primarily on atmospheric composition, incident stellar flux, and tidal evolution. Here, we develop a classification method using available data of observed stellar and planetary features, masses, and radii, to model non-thermal atmospheric escape, thermal atmospheric escape, and stellar irradiation boundaries. We compute the escape velocities for the known rocky exoplanets and compare them to the insolation and thermal velocity of likely gases to determine whether the exoplanets can maintain their atmospheres and, if so, which gases are retained. Furthermore, we quantify planetary temperature conditions based on the incident stellar fluxes to determine the likelihood of rocky exoplanets residing in a temperate or runaway greenhouse zone. We utilise these factors to classify the current list of rocky exoplanets and group them into categories based on similarity to the most appropriate solar system analogue. Subsequently, to demonstrate the functionality of this classification method, we combine it with an extension to the simple surface pressure model (Equation \ref{eq:4}), plot the adjusted surface pressure vs temperature phase diagram, and discuss the implications for the habitability of exoplanets and the optimisation of target selection for future atmospheric observations. 

\section{Method}
\subsection{Non-thermal atmospheric escape}
\label{sec:vesc}
An exoplanet must have an atmosphere to have a significant surface pressure. Thus, an important feature in determining surface pressure is the escape velocity of an exoplanet, which can be calculated as:
\begin{equation} 
\label{eq:7} 
V_{esc} = \sqrt{\frac{2GM_p}{R_p}}
\end{equation}

\noindent where $G$ is the universal gravitational constant $6.6743\times10^{-11}m^{3}kg^{-1}s^{-2}$. The escape itself is a rapid process, unlikely to be directly observable. According to \citet{zahnle2017cosmic}, the cumulative impact of escape should be evident in the statistical analysis of exoplanets, with a division between planets with and without atmospheres, that they define as the ``cosmic shoreline'' \citep{catling2009planetary,zahnle2013cosmic,zahnle2017cosmic}. The solar system planets are neatly divided around the cosmic shoreline:
\begin{equation} 
\label{eq:8} 
S_* = {5\times10^{-16}} {V_{esc}^4}
\end{equation}

\noindent where $S_*$ is insolation and $v_{esc}$ is the escape velocity as defined in Equation \ref{eq:7}. The cosmic shoreline is a simple approximation of non-thermal atmospheric escape based on observable parameters. While there are additional non-thermal processes that could increase the amount of atmospheric mass-loss, for example ion pickup \citep{lammer2006loss}, sputtering \citep{terada2009atmosphere}, dissociation and dissociative recombination \citep{geppert2008dissociative}, photo-chemical energising mechanisms \citep{vidotto2013effects}, and charge exchange \citep{dong2017dehydration}; these models require supplementary information for which we currently have no observed values and no clear method to obtain. 

\subsection{Thermal atmospheric escape}
\label{sec:tesc}
The thermal escape rate is a function of the planet's escape velocity and the temperature of the exobase. There are two types of thermal escape in atmospheres: hydrodynamic escape and slow thermal escape (Jeans escape). Highly irradiated large-mass exoplanets are more likely to lose atmospheric mass through hydrodynamic blow-off \citep{owen2019atmospheric}. However, \citet{konatham2020atmospheric} suggest that rocky exoplanets are more inclined to undergo slow thermal escape caused by atmospheric species' thermal velocities. Using the basic principles of the kinetic theory of gases, we follow \citet{konatham2020atmospheric} to predict probable atmospheric compositions of exoplanets by identifying the atmospheric species that can leave their atmospheres via slow thermal escape. According to the \citet{konatham2020atmospheric} model, the thermal escape rate is defined by the thermal velocity of the atmospheric species:
\begin{equation} 
\label{eq:9} 
U = \sqrt{\frac{3k_bT}{m}}
\end{equation}

\noindent where $m$ is the mass of a gas, $k_b$ is Boltzmann’s constant, and $T$ is the exobase temperature. As an observable value for $T$ is currently unavailable, we follow the \citet{konatham2020atmospheric} approach for a fast rotating exoplanet and utilise its equilibrium temperature, $T_{eq}$ with albedo A = 0, as a conservative approach for estimating the slow thermal escape of species from the atmosphere:
\begin{equation} 
\label{eq:11} 
T_{eq} = \left[\frac{\left(1-A\right)S_*}{4\sigma}\right]^{\frac{1}{4}}
\end{equation}

\noindent where $\sigma$ is the Stefan–Boltzmann constant. \citet{konatham2020atmospheric} infer results for rocky exoplanets experiencing slow thermal escape using data from observations of gases escaping from solar system planets' atmospheres. Equation \ref{eq:10} relates $U$ to $v_{esc}$ for atmospheric species to escape an exoplanet's atmosphere:
\begin{equation} 
\label{eq:10} 
U > \frac{1}{10}v_{esc}
\end{equation}

\subsection{Circumstellar Habitable Zone}
\label{sec:chz}
The search for potential Earth analogues begins by examining the Circumstellar Habitable Zone (CHZ), defined as the region in which a rocky planet, with favourable atmospheric conditions, can sustain liquid water on its surface \citep{kasting1993habitable,selsis2007habitable,kopparapu2013habitable,kopparapu2014habitable}. \citet{kopparapu2013habitable,kopparapu2014habitable} estimate the CHZ around stars with stellar effective temperatures ($T_*$) in the range of 2600–7200 K, for planetary masses between 0.1-5M$_\oplus$, and assuming H$_2$O (inner boundary) and CO$_2$ (outer boundary) dominated atmospheres, with N$_2$ as the background gas. This model quantifies incident stellar radiation fluxes that would result in planetary temperature conditions shifting to a runaway snowball or a runaway greenhouse. Here, we use the \citet{kopparapu2014habitable} optimistic definition of the CHZ ``early Mars'' outer (Equation \ref{eq:14}) and ``recent Venus'' inner (Equation \ref{eq:15}) boundaries:
\begin{equation} 
\label{eq:14} 
\scalebox{0.85}{$
\begin{aligned}
S_{EM} = 0.32 + 5.547\times10^{-5}(T_* - 5780) + 1.526\times10^{-9}(T_* - 5780)^2 \\
- 2.874\times10^{-12}(T_* - 5780)^3 - 5.011\times10^{-16}(T_* - 5780)^4
\end{aligned}$}
\end{equation}
\begin{equation} 
\label{eq:15} 
\scalebox{0.85}{$
\begin{aligned}
S_{RV} = 1.776 + 2.136\times10^{-4}(T_* - 5780) + 2.533\times10^{-8}(T_* - 5780)^2 \\
    - 1.332\times10^{-11}(T_* - 5780)^3 - 3.097\times10^{-15}(T_* - 5780)^4
\end{aligned}$}
\end{equation}

\noindent where $S_{EM}$ is the stellar flux required for the early Mars outer CHZ boundary, and $S_{RV}$ is the stellar flux required for the recent Venus inner CHZ boundary. These CHZ boundaries define the exoplanets in our sample that most closely resemble Earth and are therefore likely to have liquid water on their surfaces. 

\subsection{Venus Zone}
\label{sec:vz}
There is a clear distinction in atmospheric evolution between Earth and Venus, probably due to the significant difference in solar irradiance (approximately a factor of two). \citet{kane2014frequency} define a ``Venus Zone'' (VZ) where a planet is considered to be a Venus analogue rather than an Earth analogue. 

We use the optimistic CHZ ``recent Venus'' boundary from Equation \ref{eq:15} to define the runaway greenhouse outer VZ boundary, where oceans completely evaporate, resulting in the inability to execute a carbon cycle and efficiently moderate atmospheric CO$_2$ levels, leading to the formation of a thick Venus-like atmosphere. 

As distance from the host star decreases, the likelihood of substantial atmospheric mass loss increases. \citet{kane2014frequency} determine the insolation flux required for Venus to cross the \citet{zahnle2017cosmic} cosmic shoreline and use this value to denote the complete atmospheric erosion of Venus analogues ($S_{AE}$) as an approximation for the VZ inner boundary:
\begin{equation} 
\label{eq:16} 
S_{AE} \approx 25 S_\oplus
\end{equation}

Just as planets in the CHZ could be considered Earth analogues until more spectroscopic information becomes available, planets inside the VZ could be considered Venus analogues until further characterisation observations are undertaken.

\subsection{Surface pressure}
\label{sec:Psurf}
After cataloguing our sample of rocky exoplanets, we adjust parameters such as surface pressure to more appropriately normalise simple calculations. Taking the solar system planet values for surface pressure ($P_{SS-planet}$) and radius ($R_{SS-planet}$) into account, we adjust Equation \ref{eq:4} to:
\begin{equation} 
\label{eq:panalogue} 
P_{SS-analogue} = P_{SS-planet}{\left(\frac{R_p}{R_{SS-planet}}\right)}^{3.168}
\end{equation}
\begin{equation} 
\label{eq:18} 
P_{Mercury-analogue} = 1.05\times10^{-13}{R_p}^{3.168}
\end{equation}
\begin{equation} 
\label{eq:19} 
P_{Mars-analogue} = 0.0467{R_p}^{3.168}
\end{equation}
\begin{equation} 
\label{eq:20} 
P_{Venus-analogue} = 108.27{R_p}^{3.168}
\end{equation}
\begin{equation} 
\label{eq:21} 
P_{Earth-analogue} = 1.014{R_p}^{3.168}
\end{equation}

\subsection{Surface temperature}
\label{sec:Tsurf}
After calculating an adjusted $P_{surf}$ by normalising to the most appropriate solar system analogue, we apply our rocky exoplanet classification system to surface temperature estimates to plot an analogue adjusted surface pressure vs surface temperature phase diagram. Unfortunately, like surface pressure, direct measurements of surface temperature are not typically available \citep{weisfeiler2015surface} and when modelling using 3D GCMs, Earth-centric assumptions of atmospheric composition or convection depth are frequently made \citep[e.g.][]{biserud2022climate,chaverot2021background,way2017resolving,yang2014low}, creating a bias in surface temperature models. \citet{del2019albedos} correlate surface temperature ($T_{surf}$) and equilibrium temperature ($T_{eq}$), by attributing the difference between the two to a greenhouse effect:
\begin{equation} 
\label{eq:Tsurf} 
T_{surf} = T_{eq} + G_{a}
\end{equation}

\noindent where $G_a$ is the atmospheric greenhouse effect. Substituting Equation \ref{eq:11} into Equation \ref{eq:Tsurf}, we attain a surface temperature equation: 
\begin{equation} 
\label{eq:TGsurf} 
T_{surf} = \left[\frac{\left(1-A\right)S_*}{4\sigma}\right]^{\frac{1}{4}} + G_{a}
\end{equation}

Substituting observed values of the rocky solar system planets' bond albedo, insolation flux, and average surface temperature into Equation \ref{eq:TGsurf} allows us to calculate the atmospheric greenhouse effect for our solar system analogues (Table \ref{table:1}). 
   \begin{table}
      \caption[]{Solar System analogue surface temperature calculation values.}
         \label{table:1}$$ 
         \begin{tabular}{lcccc}
            \hline
            \noalign{\smallskip}
            Planet & $T_{surf}$ (K) & $A$ & $S_*$ (Wm$^{-2}$) & $G_{a}$ (K) \\
            \noalign{\smallskip}
            \hline
            \noalign{\smallskip}
            Mercury & 440 & 0.07 & 9082.7 & 0.85 \\
            Mars & 210 & 0.25 & 586.2 & -0.15\\
            Venus & 737 & 0.77 & 2601.3 & 510.85\\
            Earth & 288 & 0.3 & 1361.0 & 33.85\\
            \noalign{\smallskip}
            \hline
         \end{tabular}$$ 
   \end{table}

While we do not take the time evolution factor into account in this paper, it should be noted that the values in Table \ref{table:1} have changed over the lifetime of the solar system and similar variations are expected across the lifetime of exoplanet systems. 

This surface temperature model only accounts for a broad greenhouse effect and not specific greenhouse gas abundances. However, it enables us to apply solar system planet values for bond albedo ($A_{SS-planet}$) and atmospheric greenhouse ($G_{aSS-planet}$) from Table \ref{table:1} and refine our surface temperature calculations by adjusting Equation \ref{eq:TGsurf} to: 
\begin{equation} 
\label{eq:Tanalogue} 
T_{SS-analogue} = \left[\frac{\left(1-A_{SS-planet}\right)S_*}{4\sigma}\right]^{\frac{1}{4}} + G_{aSS-planet}
\end{equation}
\begin{equation} 
\label{eq:28} 
T_{Mercury-analogue} = 45.0 {S_*}^{\frac{1}{4}} + 0.85
\end{equation}
\begin{equation} 
\label{eq:29} 
T_{Mars-analogue} = 42.6 {S_*}^{\frac{1}{4}} - 0.15
\end{equation}
\begin{equation} 
\label{eq:30} 
T_{Venus-analogue} = 31.7 {S_*}^{\frac{1}{4}} + 510.85
\end{equation}
\begin{equation} 
\label{eq:31} 
T_{Earth-analogue} = 41.9 {S_*}^{\frac{1}{4}} + 33.85
\end{equation}

\subsection{Sample selection and Monte Carlo calculations}
\label{sec:sample}
We utilise NASA's composite planet database\footnote{\url{https://exoplanetarchive.ipac.caltech.edu/cgi-bin/TblView/nph-tblView?app=ExoTbls&config=compositepars} (accessed 18 December 2022) \label{nasa}}, in combination with additional information on the Kepler planets' radii provided by \citet{berger2020gaiaII} and updated stellar properties by \citet{berger2020gaiaI} to compose a catalogue of rocky exoplanets. Furthermore, we use the \citet{chen2016probabilistic} M-R relationship detailed in Equation \ref{eq:2} to calculate unknown radii or masses and their uncertainties. 

To ensure that we have included only rocky exoplanets in our database, we follow the \citet{chen2016probabilistic} definition of the boundary between Jovian and terrestrial planets, limiting our selection to exoplanets with radii ${R}_{p}$ $\le$ 1.23 R$_\oplus$. 

As we need information for stellar temperature, insolation flux, planetary mass, and planetary radius as provided by the NASA composite planet database, \citet{berger2020gaiaII}, and \citet{berger2020gaiaI}, our total sample contains 720 rocky exoplanets. For each exoplanet in our sample, we execute 10,000 Monte Carlo simulations using a Gaussian probability distribution for uncertainties on the stellar temperatures, insolation fluxes, planetary masses, and planetary radii as provided by the NASA composite planet database, \citet{berger2020gaiaII}, and \citet{berger2020gaiaI}. Furthermore, for the subset of exoplanets where the mass-radius relation was used, Equation \ref{eq:2} was input directly into the Monte Carlo simulation to ensure the uncertainties were appropriately correlated. 

The Monte Carlo simulations allow us to determine the median and 68\% confidence intervals on escape velocity ($v_{esc}$), equilibrium temperature ($T_{eq}$), surface temperature ($T_{surf}$), and surface pressure ($P_{surf}$). 

\section{Data analysis \& Discussion}
\subsection{Exoplanet classification}
\label{sec:catalog}
\begin{figure}
    \includegraphics[width=\columnwidth]{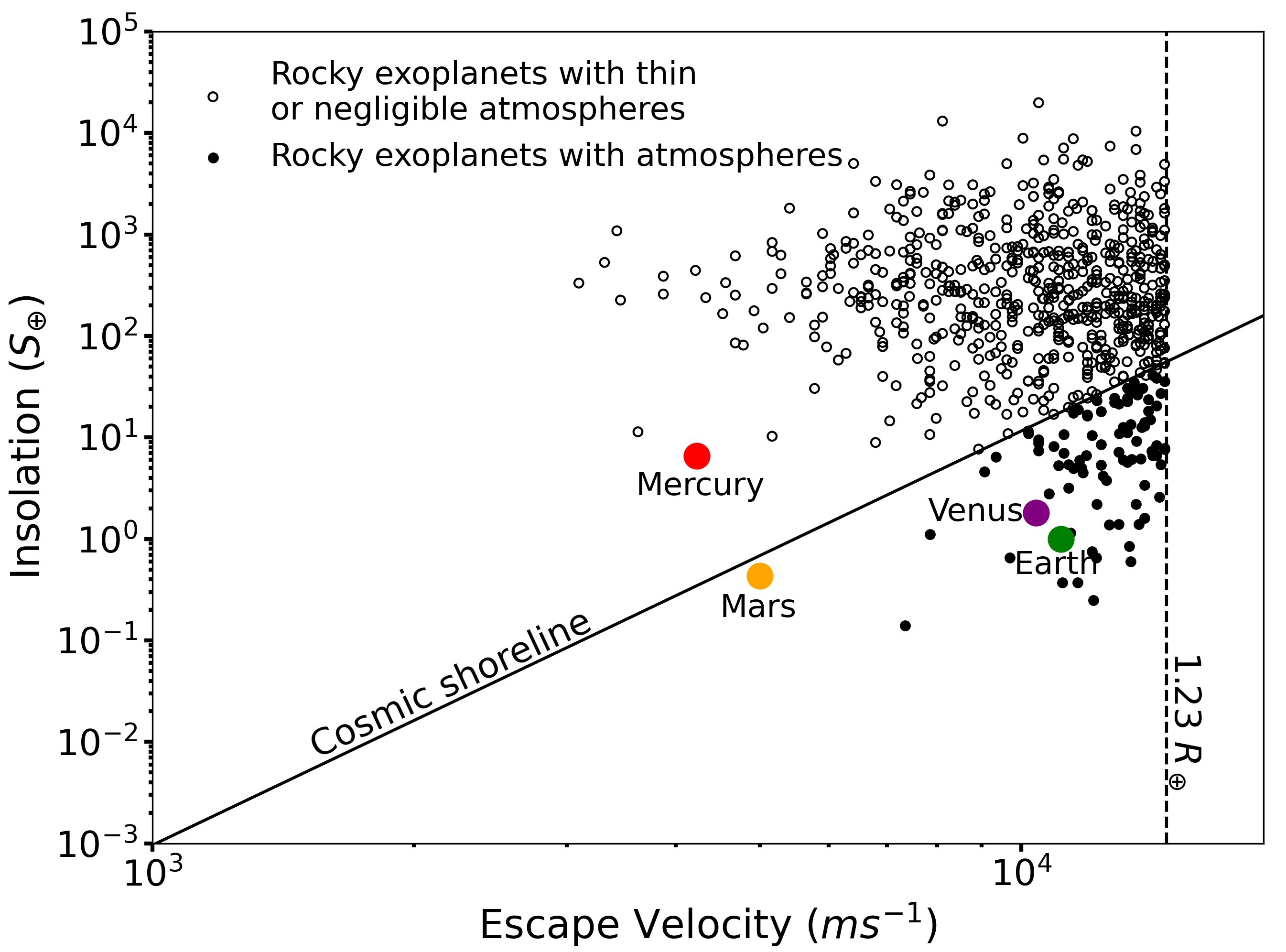}
    \caption{Insolation and escape velocity for the 720 exoplanets in our sample. Exoplanets likely to have thin or no atmosphere are plotted with open circles. Exoplanets likely to have an atmosphere are solid dots. Labelled solid coloured circles represent observed values for rocky solar system planets. The solid black line represents the \citet{zahnle2017cosmic} cosmic shoreline (Equation \ref{eq:8}). The dashed vertical line indicates the \citet{chen2016probabilistic} cut-off for rocky exoplanets, ${R}_{p}$ $\le$ 1.23 R$_\oplus$. Median and 68\% confidence intervals on escape velocity values were calculated using the Monte Carlo simulations.}
    \label{fig:Fig1}
\end{figure}

In Figure \ref{fig:Fig1}, we plot stellar insolation flux ($S_*$) against the escape velocity ($v_{esc}$) for the exoplanets in our sample as well as the four rocky solar system planets. The trend seen in Figure \ref{fig:Fig1} is what we would expect to see if escape were the primary factor influencing the volatile inventories of exoplanets. With the aid of the \citet{zahnle2017cosmic} cosmic shoreline (Equation \ref{eq:8}), we can see that planetary atmospheres are thick when the influence of the central star is weak (measured by insolation) or the gravitational well is deep (measured by escape velocity). On the other hand, we have planets with thin or no atmospheres when the star is too bright or gravity is weak. 

From our sample of 720 rocky exoplanets, 12\% $\pm$ 4\% reside below the cosmic shoreline and are consequently likely to maintain a significant atmosphere. Conversely, 88\% $\pm$ 5\% of the exoplanets in our sample reside above the cosmic shoreline and are likely to have thin or negligible atmospheres, based on the rapid escape velocity parameter alone. The uncertainty values arise from taking into account the 68\% confidence intervals on insolation and escape velocity. Taking the upper insolation error and lower escape velocity error into account, 4\% of exoplanets located in the area where rocky planets are likely to maintain a significant atmosphere, cross the cosmic shoreline and thus could potentially have thin or no atmosphere. Conversely, taking the lower insolation error and upper escape velocity error into account, 5\% of exoplanets located in the area where rocky planets have a thin or negligible atmosphere, cross the cosmic shoreline and thus could potentially have a substantial atmosphere. Furthermore, Figure \ref{fig:Fig1} indicates that most rocky exoplanets lie above the cosmic shoreline, where gravity is weak and stellar flux strong, implying that our current observational exoplanet data has a bias towards close-orbiting rocky planets with limited atmospheres. 

All rocky exoplanets discovered thus far are on close, highly irradiated orbits, bombarded by large amounts of ionising EUV and X-ray radiation \citep{lammer2022exosphere}. The upper atmosphere of an exoplanet also may be exposed to coronal mass ejections and stellar winds, inducing additional non-thermal loss processes of ion pickup \citep{lammer2006loss}, sputtering \citep{terada2009atmosphere}, dissociation and dissociative recombination \citep{geppert2008dissociative}, photo-chemical energising mechanisms \citep{vidotto2013effects}, and charge exchange \citep{dong2017dehydration}. These non-thermal escape mechanisms could shift the cosmic shoreline to lower insolation at a given escape velocity and are particularly important in the early phases of the host star’s evolution, when XUV flux and CME rates may be orders of magnitude higher \citep{do2022contribution,gronoff2020atmospheric}. However, models for these non-thermal processes require additional information for which we currently have no observed values, so it is beyond the scope of this paper. Alternatively, depending on the stellar wind pressures, a significant magnetic field could mitigate the non-thermal escape processes, resulting in a shift of the cosmic shoreline to higher escape velocities \citep{mcintyre2019planetary,egan2019planetary}. 

There are two types of thermal escape in atmospheres; hydrodynamic escape and slow thermal escape (Jeans escape). Most models have been designed for hydrodynamic conditions primarily for small semi-major axis, high-mass, highly irradiated exoplanets \citep{owen2019atmospheric,tian2015atmospheric,madhusudhan2019exoplanetary}. Conversely, low-irradiated, small-mass rocky exoplanets are more likely to experience slow thermal escape driven by thermal velocities of atmospheric species when assuming that equilibrium temperature is a good guide to thermospheric temperature \citep{konatham2020atmospheric}.

\begin{figure}
    \includegraphics[width=\columnwidth]{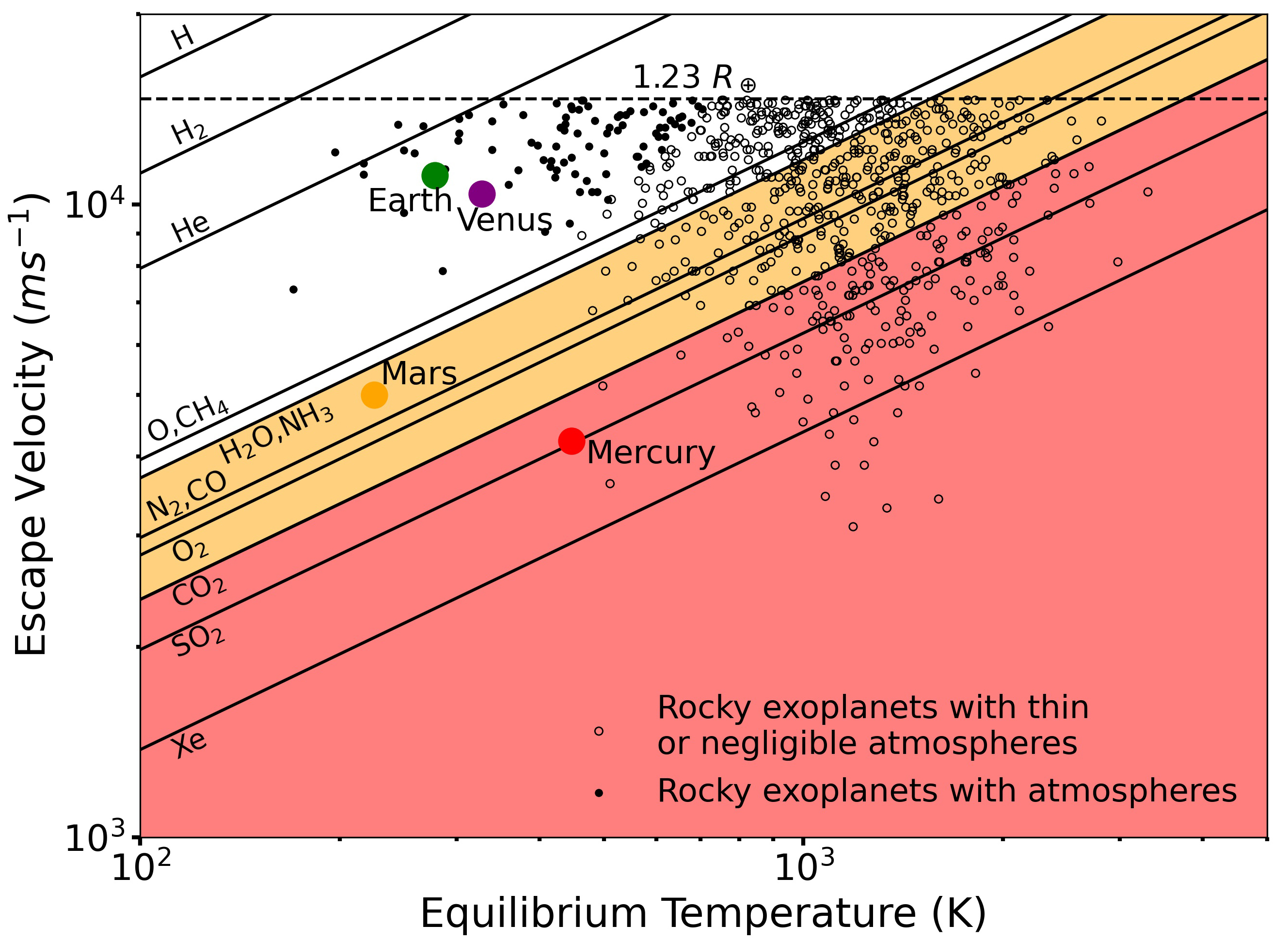}
    \caption{Escape velocity (Equation \ref{eq:7}) versus equilibrium temperature (Equation \ref{eq:11}) of our sample of 720 rocky exoplanets. Exoplanets that reside above the cosmic shoreline in Figure \ref{fig:Fig1} and are likely to have thin or no atmosphere are plotted with open circles. Exoplanets that reside below the cosmic shoreline in Figure \ref{fig:Fig1} and are likely to have an atmosphere are plotted with solid dots. Labelled solid coloured circles represent calculated values for rocky solar system planets. Solid black lines represent the thermal velocity of atmospheric gas species (defined by Equations \ref{eq:9}-\ref{eq:10}) \citep{konatham2020atmospheric}, where exoplanet atmospheres may retain the gas at lower escape velocities. Red shaded region denotes Mercury analogue exoplanets. Orange shaded region denotes Mars analogues exoplanets. The dashed horizontal line indicates the cut-off for rocky exoplanets as ${R}_{p}$ $\le$ 1.23 R$_\oplus$. Median and 68\% confidence intervals on escape velocity and equilibrium temperature values were calculated using the Monte Carlo simulations.}  
    \label{fig:Fig2}
\end{figure}

In Figure \ref{fig:Fig2}, we compute the thermal velocities of selected gases for our sample of rocky exoplanets (parameterised by the equilibrium temperature from Equation \ref{eq:11}) and compare them with the escape velocity to determine which gases could be preserved in the planets' atmospheres, in accordance with Equation \ref{eq:10}. The diagonal lines depict the thermal velocity of various atmospheric species as a function of kinetic temperature (also known as velocity lines) collated from \citet{konatham2020atmospheric}. An exoplanet can retain a specific atmospheric species if its velocity line is below the position of the exoplanet in Figure \ref{fig:Fig2}. Conversely, an atmospheric species escapes the exoplanet's atmosphere if its velocity line is above the exoplanet's position. This allows us to utilise the kinetic theory of gases to estimate potential atmospheric constituents for our sample of rocky exoplanets. 

In Figure \ref{fig:Fig2}, we can see that 22\% $\pm$ 8\% of rocky exoplanets in our sample lie below the CO$_2$ line in the red shaded region. All the exoplanets in this subset were also located substantially above the cosmic shoreline (Figure \ref{fig:Fig1}). Based on their positioning in Figures \ref{fig:Fig1} and \ref{fig:Fig2}, these exoplanets are unlikely to be able to sustain a significant atmosphere and thus will have limited surface pressure. When comparing to the four rocky solar system analogues, we can see that these exoplanets most closely resemble Mercury with its negligible atmosphere (surface pressure $\sim$5 picobars). Consequently, this subset of exoplanets could be classified as Mercury analogues.

The next group of exoplanets are those located above the CO$_2$ line and below the H$_2$O line. The 39\% $\pm$ 4\% of rocky exoplanets from our sample residing within the orange shaded region in Figure \ref{fig:Fig2} also lie above the cosmic shoreline in Figure \ref{fig:Fig1}. Based on their position in Figures \ref{fig:Fig1} and \ref{fig:Fig2}, these exoplanets are likely to have thin water-less atmospheres. When comparing to the four rocky solar system analogues, we can see that the atmospheres of these exoplanets most closely resemble present-day Mars with its thin, CO$_2$-rich, dry atmosphere (surface pressure of 0.00636 bars). Consequently, this subset of exoplanets could be classified as Mars analogues.

\begin{figure}
    \includegraphics[width=\columnwidth]{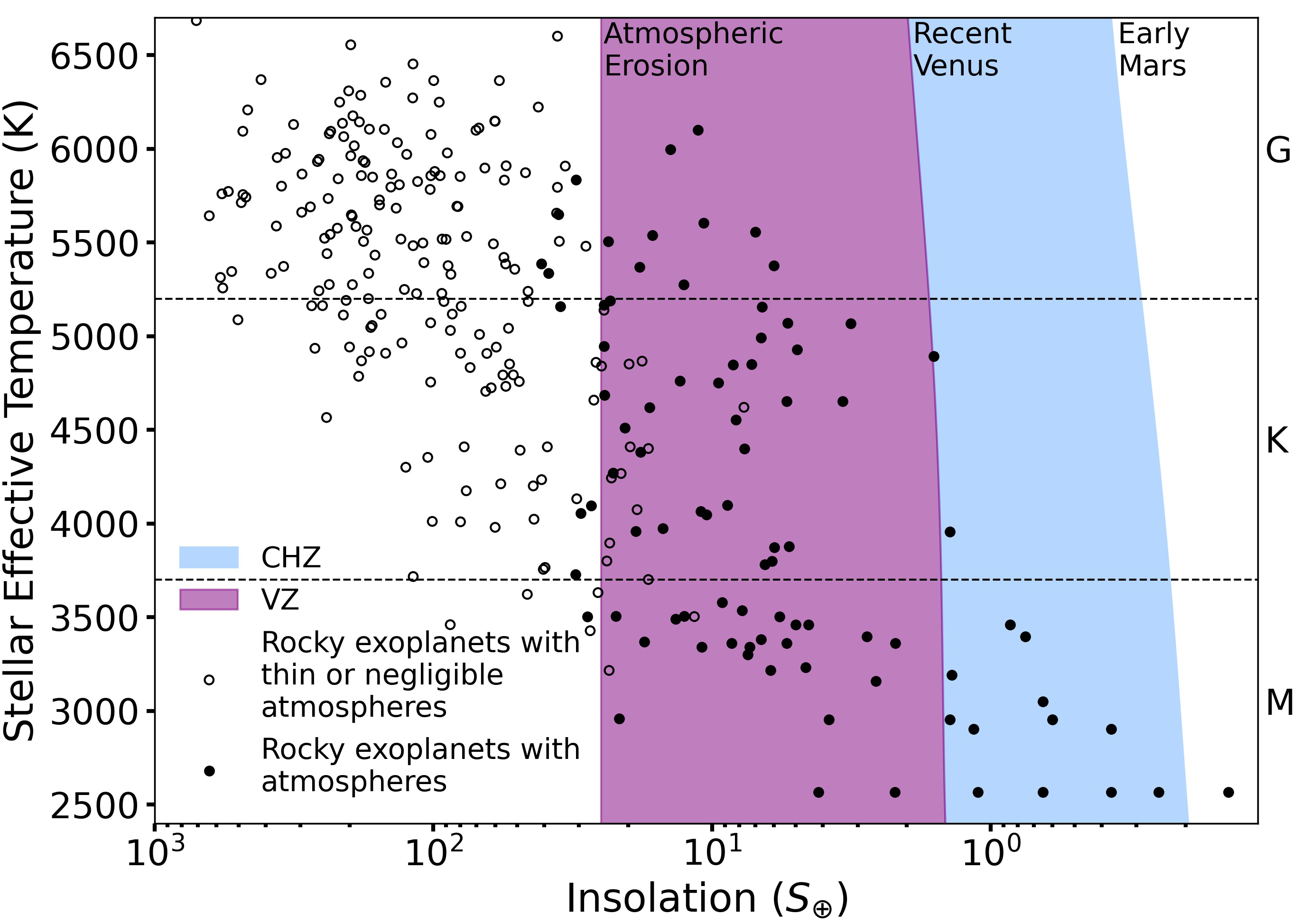}
    \caption{Location within the CHZ and VZ of 279 rocky exoplanets that reside above the H$_2$O line in Figure \ref{fig:Fig2}. The blue shaded region denotes the CHZ. The purple shaded region denotes the VZ. Exoplanets that reside above the cosmic shoreline in Figure \ref{fig:Fig1} and are likely to have thin or no atmosphere are plotted with open circles. Exoplanets that reside below the cosmic shoreline in Figure \ref{fig:Fig1} and are likely to have an atmosphere are plotted with solid dots. Trappist-1h, the exoplanet shown beyond the Early Mars boundary, is discussed further in \ref{sec:apply}}  
    \label{fig:Fig3}
\end{figure}

The final 39\% $\pm$ 12\% of rocky exoplanets presented in Figure \ref{fig:Fig2} reside above the H$_2$O line and are likely to have water in their atmospheres. When comparing to the four rocky solar system analogues, both Venus and Earth fall within similar locations in Figures \ref{fig:Fig1} and \ref{fig:Fig2} yet have substantially different atmospheres with surface pressures differing by two orders of magnitude. To determine which of the 279 rocky exoplanets located above the H$_2$O line in Figure \ref{fig:Fig2} are likely to be Earth analogues or Venus analogues, we examine their location in relation to the CHZ and VZ in Figure \ref{fig:Fig3}. 

Using Equations \ref{eq:14} - \ref{eq:15}, we plot the optimistic boundaries of the CHZ in Figure \ref{fig:Fig3} to quantify the insolation flux thresholds where planetary conditions transition to either a runaway snowball or a runaway greenhouse. The 2\% $\pm$ 1\% of rocky exoplanets that are likely to have an atmosphere (Figure \ref{fig:Fig1}), with H$_2$O present (Figure \ref{fig:Fig2}), and also reside within the optimistic CHZ (Figure \ref{fig:Fig3}) where the insolation is suitable for the presence of liquid water on the planets' surfaces, are the planets most similar to Earth. Thus, these exoplanets could be classified as Earth analogues and are the most likely to retain a $\sim$1 bar atmosphere including H$_2$O molecules, indicating a high potential for retaining water in their atmospheres and on their surfaces. 

Additionally, using Equations \ref{eq:15} - \ref{eq:16}, we plot the boundaries of the VZ in Figure \ref{fig:Fig3} to quantify the insolation flux thresholds where planetary conditions transition between a runaway greenhouse and complete atmospheric erosion. In Figure \ref{fig:Fig3}, we see that 11\% $\pm$ 1\% of rocky exoplanets with H$_2$O present in their atmospheres (Figure \ref{fig:Fig2}) reside within the VZ. Out of these 11\% $\pm$ 1\% rocky exoplanets, fifteen are plotted with open circles denoting their location above the cosmic shoreline in Figure \ref{fig:Fig1}, indicating they are unlikely to have an atmosphere. However, taking into account 68\% confidence intervals on insolation and escape velocity, these exoplanets cross the cosmic shoreline and could potentially have a substantial atmosphere. Thus, the 11\% $\pm$ 1\% of rocky exoplanets within uncertainties are likely to have an atmosphere (Figure \ref{fig:Fig1}), with H$_2$O present (Figure \ref{fig:Fig2}). Their position within the VZ (Figure \ref{fig:Fig3}), indicates that their atmospheres would be unable to maintain radiation balance, resulting in runaway heating of the surface and the formation of a thick Venus-like atmosphere and surface pressures in the order of $10^2$ bar. Therefore, this subset of rocky exoplanets could be classified as Venus analogues. 

The final subset of 26\% $\pm$ 12\% rocky exoplanets from our sample reside above the H$_2$O line in Figure \ref{fig:Fig2} and are not located within the CHZ or VZ in Figure \ref{fig:Fig3}. Furthermore, the majority of planets in this subset are likely to have thin or negligible atmospheres, according to Figure \ref{fig:Fig1}. Out of these 26\% $\pm$ 12\% rocky exoplanets, nine exoplanets are located past the inner VZ atmospheric erosion boundary, yet they are plotted with solid circles, indicating they are likely to have an atmosphere according to Figure \ref{fig:Fig1}. However, taking into account 68\% confidence intervals on insolation and escape velocity, these exoplanets cross the cosmic shoreline and could potentially have a thin or negligible atmosphere. \citet{kane2014frequency} suggest that these highly irradiated rocky exoplanets located past the VZ's inner boundary have completely eroded atmospheres. No solar system analogues exist for this subset of rocky exoplanets, which have the potential for H$_2$O to be present, yet too high levels of stellar flux eroding their atmospheres. 

\begin{figure*}
    \includegraphics[width=\textwidth]{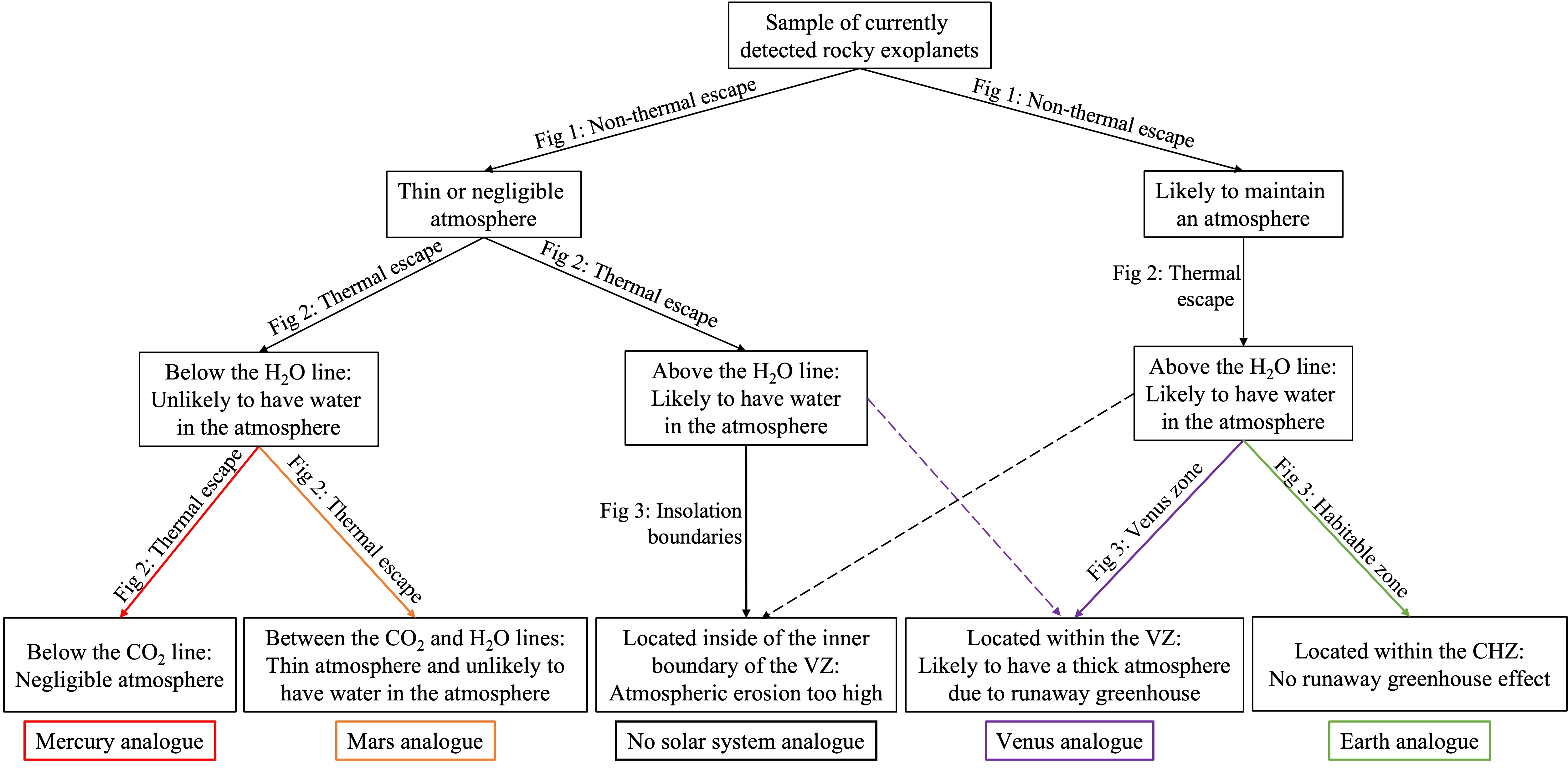}
    \caption{Classification of rocky exoplanets into closest solar system analogue. The dashed lines indicate potential for crossover in classification due to uncertainties in the input parameters. None of the exoplanets that were likely to maintain an atmosphere were below the H$_2$O line.}
    \label{fig:Fig5}
\end{figure*}

Combining the information from Figures \ref{fig:Fig1}-\ref{fig:Fig3}, we develop the classification method outlined in Figure \ref{fig:Fig5}. Comparing to previous classification schemes, \citet{wordsworth2022atmospheres} classify a subset of rocky exoplanets that have an atmosphere and equilibrium temperatures above 300K as having no direct analogue in the solar system. However, here, we classify these exoplanets as Venus-like as we have determined that they retain an atmosphere that could have H$_2$O present despite their high temperatures. This subset will be interesting to explore in future observations to study the key transitions in atmospheric composition and determine how they differ from Venus. Additionally, in Figure 3 we have classified a different subset than \citet{wordsworth2022atmospheres} having no solar system analogue as those rocky exoplanets that have the potential for H$_2$O to be present, yet due to high levels of stellar flux, have no or negligible atmospheres.

\subsection{Application of exoplanet classification to surface pressure and surface temperature}
\label{sec:apply}

\begin{figure*}
	\includegraphics[width=\textwidth]{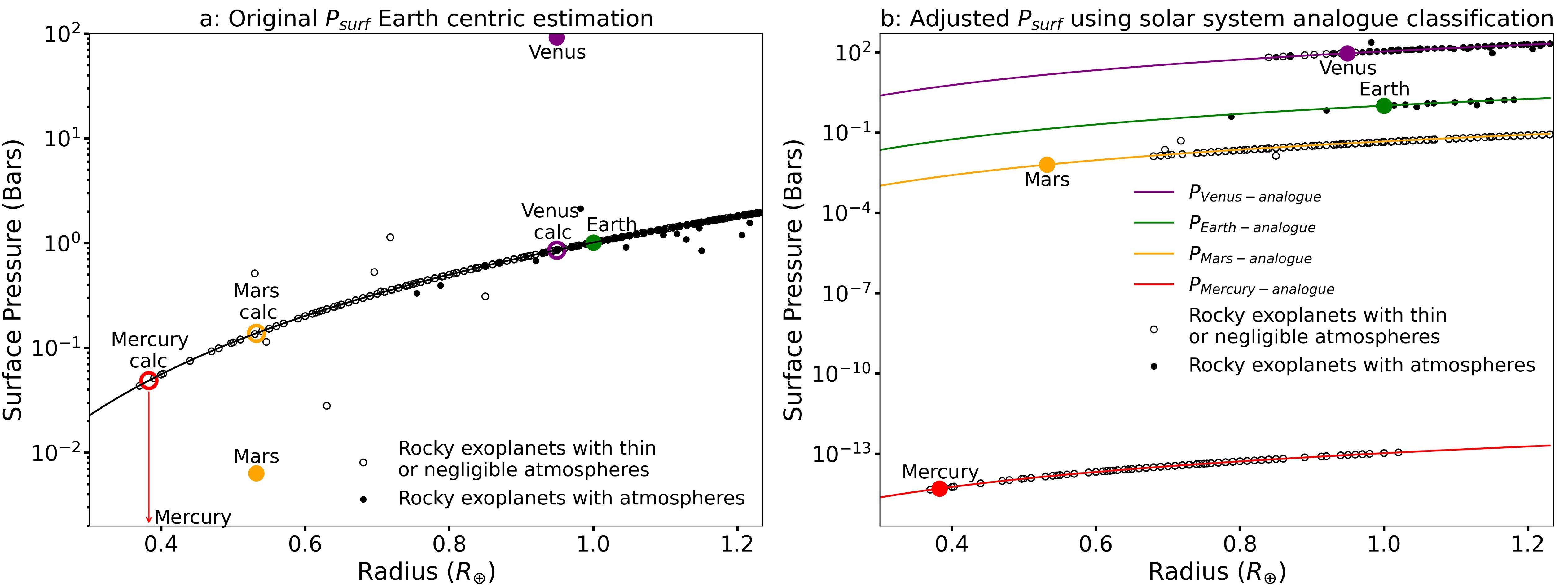}
    \caption{\textbf{a.} Surface pressure calculation using original Equation \ref{eq:4} applied to 720 rocky exoplanets in our sample. Open coloured circles represent the calculated values for rocky solar system planets. Labelled solid coloured circles represent observed values for rocky solar system planets. Exoplanets that reside above the cosmic shoreline in Figure \ref{fig:Fig1} and are likely to have thin or no atmosphere are plotted with open circles. Exoplanets that reside below the cosmic shoreline in Figure \ref{fig:Fig1} and are likely to have an atmosphere are plotted with solid dots. Median and 68\% confidence intervals on surface pressure values were calculated using the Monte Carlo simulations. \textbf{b.} Surface pressure calculation using Equations \ref{eq:18}-\ref{eq:21} applied to our sample categorised according to Figure \ref{fig:Fig5}. Labelled solid coloured circles represent observed values for rocky solar system planets. Exoplanets that reside above the cosmic shoreline in Figure \ref{fig:Fig1} and are likely to have thin or no atmosphere are plotted with open circles. Exoplanets that reside below the cosmic shoreline in Figure \ref{fig:Fig1} and are likely to have an atmosphere are plotted with solid dots. Exoplanets that are classified as having ``no solar system analogue'' have no analogue adjusted equations and are thus omitted from this graph. Median and 68\% confidence intervals on surface pressure values were calculated using the Monte Carlo simulations.}  
    \label{fig:Fig6}
\end{figure*}

To demonstrate the effect that the new classification system has on simple normalised calculations, the surface pressures for all 720 rocky exoplanets in our sample were computed using the original Equation \ref{eq:4} (Figure \ref{fig:Fig6}a), and using the new Equations \ref{eq:18}-\ref{eq:21} (Figure \ref{fig:Fig6}b). 

Figure \ref{fig:Fig6}a demonstrates the clustering around $\sim$1 bar surface pressure due to the fact that surface pressure was normalised by Earth’s value of 1.014 bar. Therefore, relative to their observed values, the calculated values for Mercury and Mars are higher, and Venus is lower. Furthermore, as we have limited the radius of rocky exoplanets in our sample to $0.3R_\oplus \le R_p \le 1.23R_\oplus$, the surface pressure range is limited to $0.02-1.95$ bar. 

Figure \ref{fig:Fig6}b highlights the effect that the solar system analogue classification method has made to the surface pressure model. There is now a broader spread of surface pressure values for rocky exoplanets ranging from $2\times10^{-15} - 210$ bars. Additionally, Figure \ref{fig:Fig6}b illustrates the division between planets that have thin to no atmospheres being Mercury or Mars analogues and planets with atmospheres being Venus or Earth analogues. The exceptions to this division are the Venus analogue exoplanets whose median escape velocities indicate an absence of an atmosphere, however within 68\% confidence intervals, these Venus analogues cross the cosmic shoreline from Figure \ref{fig:Fig1} and indicate the potential for a thick Venus like atmosphere, based on their location in Figure \ref{fig:Fig3}. Exoplanets that are classified as having ``no solar system analogue'' have no analogue adjusted equations and are thus omitted from Figure \ref{fig:Fig6}b. 

\begin{figure*}
	\includegraphics[width=\textwidth]{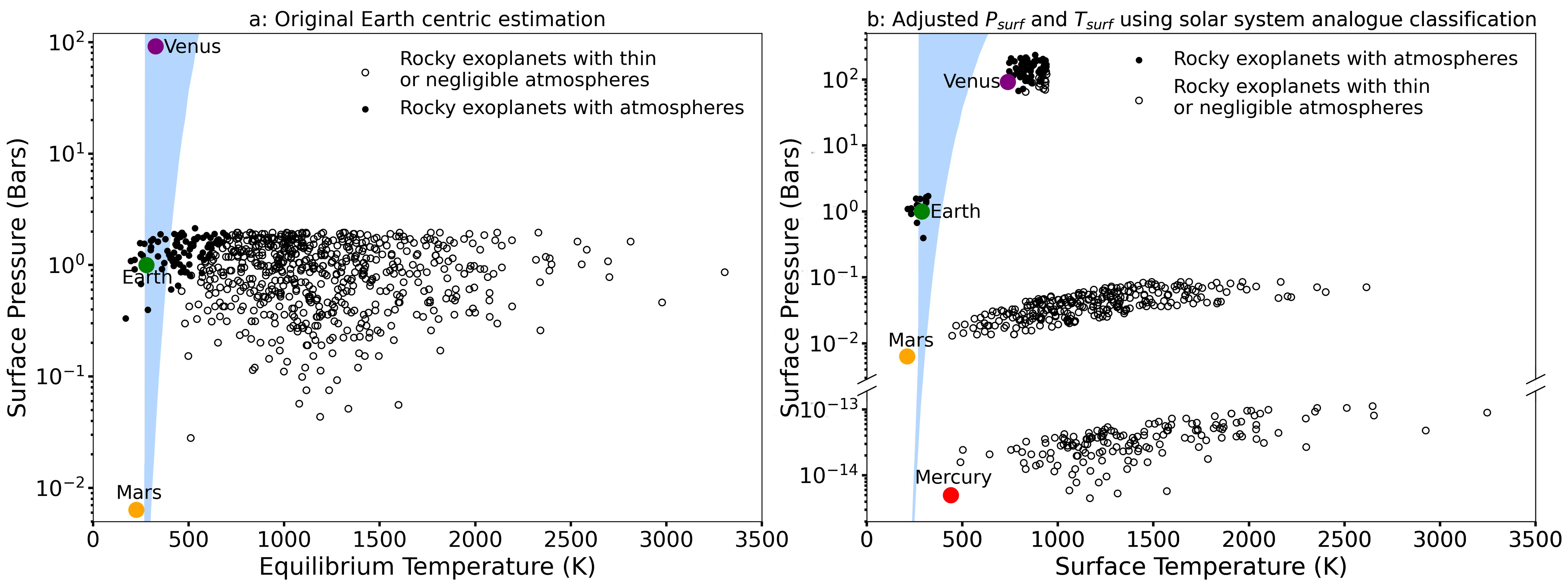}
    \caption{\textbf{a.} Phase diagram of pure water (blue shaded region) with surface pressure vs equilibrium temperature, applying Equations \ref{eq:4} and \ref{eq:11} to the 720 rocky exoplanets in our sample. Labelled solid coloured circles represent rocky solar system planets. Exoplanets that reside above the cosmic shoreline in Figure \ref{fig:Fig1} and are likely to have thin or no atmosphere are plotted with open circles. Exoplanets that reside below the cosmic shoreline in Figure \ref{fig:Fig1} and are likely to have an atmosphere are plotted with solid dots. Median and 68\% confidence intervals on surface temperature and surface pressure values were calculated using the Monte Carlo simulations. \textbf{b.} Phase diagram of pure water (blue shaded region) with adjusted surface pressure vs surface temperature, applying Equations \ref{eq:18}-\ref{eq:21} and \ref{eq:28}-\ref{eq:31} to our sample classified according to Figure \ref{fig:Fig5}. Labelled solid coloured circles represent rocky solar system planets. Exoplanets that reside above the cosmic shoreline in Figure \ref{fig:Fig1} and are likely to have thin or no atmosphere are plotted with open circles. Exoplanets that reside below the cosmic shoreline in Figure \ref{fig:Fig1} and are likely to have an atmosphere are plotted with solid dots. Exoplanets that are classified as having ``no solar system analogue'' have no analogue adjusted equations and are thus omitted from this graph. Median and 68\% confidence intervals on surface temperature and surface pressure values were calculated using the Monte Carlo simulations.}  
    \label{fig:Fig7}
\end{figure*}

Reliable estimates of surface temperature and pressure are essential for characterising the environment and habitability of an exoplanet. To further demonstrate the functionality of our exoplanet classification method, we plot our sample of rocky exoplanets over the phase diagram of pure water. In Figure \ref{fig:Fig7}a we plot the equilibrium temperature, where $A=0$ and $G_a=0$ (Equation \ref{eq:11}), against the original surface pressure values (Equation \ref{eq:4}). Subsequently, after applying our solar system analogue classification, in \ref{fig:Fig7}b we plot the surface temperature, with analogue-defined $A$ and $G_a$ values (Equations \ref{eq:28}-\ref{eq:31}), against the analogue-adjusted surface pressure values (Equations \ref{eq:18}-\ref{eq:21}). 

Figure \ref{fig:Fig7}a illustrates the 0.02-1.95 bar limitations and clustering around $\sim$1 bar surface pressure due to the Earth-centric normalisation in Equation \ref{eq:4}. Additionally, the equilibrium temperature of Venus, without accounting for any atmospheric greenhouse effect, falsely indicates the likelihood of liquid water on Venus's surface. Thus, our subset of Venus analogue exoplanets which reside within the liquid water zone in Figure \ref{fig:Fig7}a may display similar false-positive results.

Figure \ref{fig:Fig7}b highlights the effect that the solar system analogue classification method has made to the surface temperature vs pressure phase diagram. The Venus analogues have shifted outside of the liquid water zone, representing the significant effect an atmospheric greenhouse plays in the potential habitability of an exoplanet. In Figure \ref{fig:Fig7}a there are no rocky exoplanets with an atmosphere recording equilibrium temperature above 620K, however with our analogue adjusted surface temperatures in Figure \ref{fig:Fig7}b, we see some Venus-like exoplanets with surface temperatures of up to 950K, which could be classified as \textit{Atmosphere type I} from \citet{miguel2011compositions}. On these hot rocky exoplanets, the major gases present are likely to be Na, O{$_2$}, O, and Fe, with the near-crust atmospheres mainly composed of H$_2$O, CO$_2$, and SO$_2$ \citep{herbort2020atmospheres,miguel2011compositions,miguel2013hot,schaefer2009chemistry,schaefer2012vaporization}.

Furthermore, in Figure \ref{fig:Fig7}b there are six Earth analogues that are likely to have an atmosphere (Figure \ref{fig:Fig1}) with H$_2$O present (Figure \ref{fig:Fig2}) and reside within the optimistic CHZ (Figure \ref{fig:Fig3}), yet are located to the left of the liquid water zone in Figure \ref{fig:Fig7}b indicating a higher potential for water molecules to be in ice form. 

\begin{figure}
    \includegraphics[width=\columnwidth]{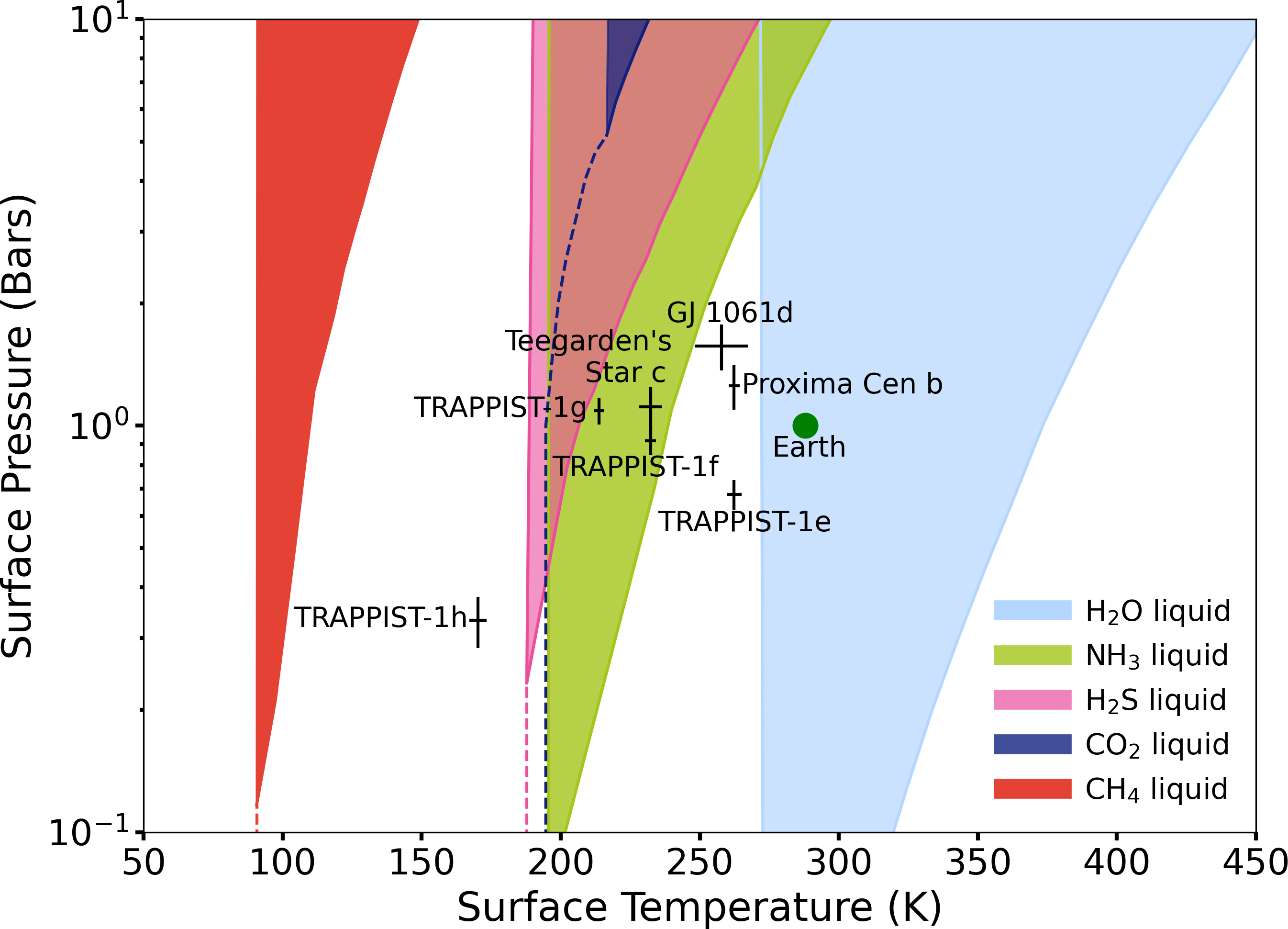}
    \caption{Analogue adjusted surface pressure and surface temperature of GJ 1061 d, Proxima Cen b, Teegarden's Star c, TRAPPIST-1 e, TRAPPIST-1 f, TRAPPIST-1 g, and TRAPPIST-1 h in the context of liquid stability for H$_2$O (light blue shading), NH$_3$ (green shading), H$_2$S (pink shading), CO$_2$ (dark blue shading), and CH$_4$ (red shading) adapted from \citet{cengel2012ebook}. The dashed coloured lines for CH$_4$ (red dashed line), H$_2$S (pink dashed line), and CO$_2$ (dark blue dashed line), represent the solid-gas boundaries. Median and 68\% confidence intervals on analogue adjusted surface pressure and surface temperature values were calculated using the Monte Carlo simulations.}
    \label{fig:Fig8}
\end{figure}

To determine the types of ices likely present on the surface of the six Earth analogues located in the water ice zone in Figure \ref{fig:Fig7}b, we plot their analogue adjusted surface pressure and surface temperature values in the context of phase diagrams for H$_2$O, NH$_3$, H$_2$S, CO$_2$, and CH$_4$ in Figure \ref{fig:Fig8} \citep{cengel2012ebook}. From Figure \ref{fig:Fig8} it is evident that all of the planets are likely to form H$_2$O ices. The potential addition of surface water ice on these exoplanets could increase their albedos, resulting in a corresponding decrease to their surface temperatures from Equation \ref{eq:TGsurf}. As more atmospheric gas is trapped in ice form, this could also result in a decrease of atmospheric pressure. Additionally, Teegarden's Star c, TRAPPIST-1 f, and TRAPPIST-1 g fall within the liquid NH$_3$ phase, indicating that if ammonia is present on their surfaces it is likely to be in liquid form. All of these exoplanets will still maintain H$_2$S, CO$_2$, and CH$_4$ in gas form in their atmospheres.

TRAPPIST-1 h is the only rocky exoplanet in our sample that resides past the Early Mars CHZ outer boundary in Figure \ref{fig:Fig3}, and is the left-most exoplanet in Figure \ref{fig:Fig7}a. Based on its location in Figures \ref{fig:Fig1} and \ref{fig:Fig2}, TRAPPIST-1 h is likely to retain an atmosphere with H$_2$O present. However, due to its location in Figures \ref{fig:Fig3} and \ref{fig:Fig7}a, TRAPPIST-1 h is likely to be cooler than the Earth analogues, and there is a higher potential for the molecules to be in ice form. From Figure \ref{fig:Fig8} it is evident that TRAPPIST-1 h is likely to form H$_2$O, NH$_3$, H$_2$S, and CO$_2$ ices, while CH$_4$ is likely to remain in gas form. 

As evident in Figures \ref{fig:Fig6} and \ref{fig:Fig7}, our rocky exoplanet classification, when applied to surface pressure and surface temperature models, allows us to present a more appropriate picture of the current rocky exoplanet sample and provides a further layer of context in the characterisation of exoplanets. 

Our model predictions can be tested using two rocky exoplanets for which the upper limits on atmospheric features have been determined using secondary eclipse and phase variation observations. First, we compared our results to the \citet{kreidberg2019absence} observations of LHS 3844b. While LHS 3844b’s radius measurement of 1.3R$_\oplus$ is technically outside our rocky exoplanet cut-off range (${R}_{p}$ $\le$ 1.23R$_\oplus$), when run through our models we find results consistent with \citet{kreidberg2019absence} of a hot rocky planet unable to retain a substantial atmosphere. Using our classification method, we categorise LHS 3844b as a “Non-solar system analogue” exoplanet as it has the potential for H$_2$O to be present, yet high levels of stellar flux have likely eroded the atmosphere. This is consistent with the findings from \citet{kreidberg2019absence} of a small atmosphere susceptible to erosion by stellar winds and thus likely being bare-rock with low bond-albedo. Additionally, we compared our results to the \citet{crossfield2022gj} observations of GJ 1252b. While our model assumes a fast-rotating exoplanet, unlike the slow-rotating GJ 1252b, when run through our models, the results agree with \citet{crossfield2022gj} that GJ 1252b is a hot rocky exoplanet with no significant atmosphere. Using our classification method, we categorise GJ 1252b as a close-in “Mars-analogue” exoplanet likely to have a thin water-less atmosphere. Comparing our analogue adjusted temperature of 1011 $\pm$ 107 K to the \citet{crossfield2022gj} dayside brightness temperature of $1410${\raisebox{0.5ex}{\tiny$^{+91}_{-125}$}} K, we fall within ~2$\sigma$ of their dayside GJ 1252b surface temperature. Despite using different assumptions, our primary classification model is consistent with other work using different techniques, giving us confidence in our approach.

\subsection{Model limitations}
\label{sec:limits}
The modelling employed in this paper provides a general approximation of atmospheric escape velocity, thermal escape of species from the atmosphere, and stellar irradiation boundaries for known exoplanets; however, it is important to acknowledge its simplicity. 

The \citet{zahnle2017cosmic} cosmic shoreline definition assumes an average molecular weight of the atmospheric material to represent all atmospheric compositions. However, if the atmospheric material is composed of significantly heavier or lighter elements, or if photochemistry which could significantly reduce the mean molecular mass is considered, then the cosmic shoreline's ideal gas assumption may be under- or over-estimating atmospheric escape rates. Furthermore, the time-frame for atmospheric loss will differ depending on luminosity and stellar mass for a given planet. Thermal loss rates may also be influenced by atmospheric composition, as \citet{parkinson2022venus} show that an O-CO$_2$ thermosphere can cool efficiently. Atmospheric species can be lost through several additional non-thermal processes for example ion pickup \citep{lammer2006loss}, sputtering \citep{terada2009atmosphere}, dissociation and dissociative recombination \citep{geppert2008dissociative}, photo-chemical energising mechanisms \citep{vidotto2013effects}, and charge exchange \citep{dong2017dehydration}. Conversely, these atmospheric species could be gained through volcanic degassing \citep{oosterloo2021role} or impact events that liberate gas from the surface \citep{kuwahara2015molecular}. Furthermore, depending on the stellar wind pressures, the presence of a significant magnetic field could reduce non-thermal atmospheric erosion \citep{mcintyre2019planetary, egan2019planetary}. Future research should be conducted to investigate which atmospheric mass-loss processes dominate in different planetary scenarios and the degree to which they are inhibited by potential planetary magnetism. 

Planetary rotation is an essential factor that could affect the modelling employed here; however, this parameter is not yet able to be directly observed for a broad sample of exoplanets. \citet{yang2014low} demonstrate the dependence of the inner CHZ boundary on planetary rotation. Strongly irradiated, rapidly rotating exoplanets could lose water and develop low albedos, entering a runaway greenhouse even while residing within the CHZ \citep{del2019albedos,Kopparapu2016inner}. Furthermore, species in the atmospheres of a rapidly rotating exoplanet could reach high velocities due to the increased temperature, which would accelerate the atmospheric escape process \citep{konatham2020atmospheric}. While we can infer the rotation period for tidally locked exoplanets as equivalent to their orbital period, such planets could escape synchronous rotation by being captured in spin-orbit resonances \citep{goldreich1966q,makarov2012dynamical,rodriguez2012spin} or through resonant planet-planet interactions with their exterior planetary companions \citep{delisle2017spin,vinson2017spin,zanazzi2019ability}. Additionally, there is no way to constrain the rotation period without direct observations of exoplanets past the tidal locking radius. In the future, photometric variability techniques have been postulated to facilitate measurements of an exoplanet’s rotation \citep{fujii2012mapping,snellen2014fast}.

Here, we only focus on the impact of stellar flux on the runaway greenhouse boundary, as defined by \citet{kopparapu2014habitable}. However, there are additional ways surface temperature conditions could increase to levels resulting in a runaway greenhouse, such as tidal heating \citep{barnes2013tidal,mcintyre2022tidally}, especially for planets in eccentric orbits \citep{williams2002earth,kane2012habitable}, or sufficiently increased CO$_2$ levels that could drive an atmosphere into a runaway greenhouse at further distances from the host star \citep{kane2014frequency}. Nonetheless, the likelihood of a runaway greenhouse will decrease dramatically as the distance past the current VZ increases. 

The surface temperature estimate utilised here only accounts for a broad greenhouse effect and not specific greenhouse gas abundances. For synchronously rotating exoplanets, future observations of thermal phase curves could help further quantify the difference between surface temperature and equilibrium temperature and provide a more accurate picture of an exoplanet's atmospheric greenhouse effect \citep{del2019habitable,lacis2010atmospheric,yang2014low}. It would also be useful to measure exoplanets' obliquity, as this can result in high to low temperatures distributed from the equator to the poles \citep{nowajewski2018atmospheric}. For example, \citet{wang2016effects} suggests that with higher obliquities, the habitability around M dwarfs narrows. Obliquity is currently unobservable for exoplanets, although it could be extrapolated when seasonal cycle information on reflected starlight becomes available \citep{kane2017obliquity}. Furthermore, \citet{ahlers2016gravity} determine that exoplanets with inclined orbits around fast-rotating stars display changes in equilibrium temperature of up to 15\% due to the increased irradiance near the stellar poles. 

Through these simplified models, we are attempting to use the limited observational data on the characteristics of rocky exoplanets we currently have available. In the future, additional observations of obliquity, inclination, planetary rotation rates, stellar rotation rates, and thermal phase curves could help strengthen the classification method detailed here and help determine optimum targets for future atmospheric observations of rocky exoplanets.

\section{Conclusions}
Here, we use non-thermal atmospheric escape, thermal atmospheric escape, and stellar irradiation boundaries to develop a primary classification method for current rocky exoplanets (${R}_{p}$ $\le$ 1.23R$_\oplus$) and group them into categories relative to the most appropriate solar system analogue. When applying this primary classification method to the 720 rocky exoplanets in our sample, results suggest that 22\% $\pm$ 8\% are Mercury analogues, 39\% $\pm$ 4\% are Mars analogues, 11\% $\pm$ 1\% are Venus analogues, 2\% $\pm$ 1\% are Earth analogues, and 26\% $\pm$ 12\% are without a known planetary counterpart in our solar system. 

Implementing this classification method will help further characterise the detected exoplanets by comparing them to a more appropriate solar system analogue, rather than continuing with the common approach where we estimate the values for numerous unknown parameters by normalising to Earth. To demonstrate the functionality of this classification method, we compare it to a simple model for surface pressure. Using Earth-centric measurements the calculated surface pressure range for rocky exoplanets ($0.3R_\oplus \le R_p \le 1.23R_\oplus$) with the simple model spanned $0.02-1.95$ bars. After applying the new primary classification method to the rocky exoplanet sample, the surface pressure range now expands to $2\times10^{-15} - 210$ bars, accounting for the full variation observed in our solar system. Furthermore, our new rocky exoplanet classification method, when applied to calculating surface pressure and surface temperature, allows us to present a more varied picture of the current rocky exoplanet sample and provides a further layer of context in the characterisation of exoplanets; for example, the presence of liquid, gas or ice.

The use of our new primary classification method could improve inferences of temperature, composition, interior structure, evolution and dynamics of rocky exoplanets, which would aid our ability to interpret and model exoplanet atmospheres \citep{kane2019venus}. Additionally, this classification method could benefit target selection for exoplanet characterisation missions by providing a more robust starting point for potential atmospheric properties and composition for comparison to future observations. 

\section*{Acknowledgements}

S.R.N.McIntyre gratefully acknowledges an Australian Government Research Training Program (RTP) Scholarship. P.L. King acknowledges funding from an Australian Research Council Discovery Program grant (DP200100406). Helpful comments from an anonymous referee are gratefully acknowledged.

\section*{Data Availability}

The data underlying this article are available in the article and in its online supplementary material which can be downloaded in electronic form from the Centre de Données astronomiques de Strasbourg (CDS) service via anonymous ftp \url{cdsarc.u-strasbg.fr} (130.79.128.5) or via 
\url{https://cdsarc.cds.unistra.fr/viz-bin/cat/J/MNRAS}.



\bibliographystyle{mnras}
\bibliography{srnmcsp} 




\appendix
\onecolumn
\section{Classification of Rocky Exoplanet Sample}
\label{appendix:a}

\begin{table}
\tiny
\begin{longtable}{lccccccccc}
\caption{TRAPPIST-1 system as an excerpt of the rocky exoplanet classification method catalogue. The rest of the table can be downloaded in electronic form from CDS service and the publisher’s website.}\\
\label{tab:1}\\
\hline\hline
Planet Name & M$_p$ (M$_\oplus$) & R$_p$ (R$_\oplus$) & v$_{esc}$ (ms$^{-1}$) & S$_*$ (S${_\oplus}$) & T$_{eq}$ (K) & Original P$_{surf}$ (Bar) & Adjusted P$_{surf}$ (Bar) & Adjusted T$_{surf}$ (K) & Planet Classification\\
\hline\hline
\endfirsthead
\caption{Continued.} \\
\hline\hline
Planet Name & M$_p$ (M$_\oplus$) & R$_p$ (R$_\oplus$) & v$_{esc}$ (ms$^{-1}$) & S$_*$ (S${_\oplus}$) & T$_{eq}$ (K) & Original P$_{surf}$ (Bar) & Adjusted P$_{surf}$ (Bar) & Adjusted T$_{surf}$ (K) & Planet Classification\\
\hline\hline
\endhead 
\hline
\endfoot
\hline
\endlastfoot
TRAPPIST-1 b & 1.37 $\pm$ 0.07 & 1.12 $\pm$ 0.01 & 12411.61 $\pm$ 322.28 & 4.15 $\pm$ 0.16 & 397.26 $\pm$ 3.82 & 1.23 $\pm$ 0.14 & 136.63 $\pm$ 08.53 & 785.66 $\pm$ 2.65 & Venus analogue \\
TRAPPIST-1 c & 1.31 $\pm$ 0.06 & 1.10 $\pm$ 0.01 & 12214.27 $\pm$ 270.09 & 2.21 $\pm$ 0.09 & 339.36 $\pm$ 3.44 & 1.20 $\pm$ 0.12 & 132.63 $\pm$ 08.43 & 745.61 $\pm$ 2.38 & Venus analogue \\
TRAPPIST-1 d & 0.39 $\pm$ 0.01 & 0.79 $\pm$ 0.01 & 7849.09 $\pm$ 131.55 & 1.11 $\pm$ 0.04 & 285.69 $\pm$ 2.60 & 0.40 $\pm$ 0.03 & 0.40 $\pm$ 0.03 & 295.07 $\pm$ 2.35 & Earth analogue \\
TRAPPIST-1 e & 0.69 $\pm$ 0.02 & 0.92 $\pm$ 0.01 & 9701.21 $\pm$ 166.83 & 0.65 $\pm$ 0.03 & 249.91 $\pm$ 2.86 & 0.68 $\pm$ 0.06 & 0.68 $\pm$ 0.06 & 262.36 $\pm$ 2.63 & Earth analogue \\
TRAPPIST-1 f & 1.04 $\pm$ 0.03 & 1.05 $\pm$ 0.01 & 11153.63 $\pm$ 180.93 & 0.37 $\pm$ 0.01 & 217.08 $\pm$ 1.46 & 0.92 $\pm$ 0.07 & 0.92 $\pm$ 0.07 & 232.34 $\pm$ 1.35 & Earth analogue \\
TRAPPIST-1 g & 1.32 $\pm$ 0.04 & 1.13 $\pm$ 0.01 & 12099.61 $\pm$ 186.13 & 0.25 $\pm$ 0.01 & 196.81 $\pm$ 1.99 & 1.09 $\pm$ 0.08 & 1.09 $\pm$ 0.08 & 213.80 $\pm$ 1.83 & Earth analogue \\
TRAPPIST-1 h & 0.33 $\pm$ 0.02 & 0.76 $\pm$ 0.01 & 7350.25 $\pm$ 234.95 & 0.14 $\pm$ 0.01 & 170.25 $\pm$ 3.03 & 0.33 $\pm$ 0.05 & No solar system analogue & No solar system analogue & No solar system analogue \\
\hline
\end{longtable}
\end{table}

\bsp	
\label{lastpage}
\end{document}